\documentclass[runningheads]{llncs}
\usepackage{graphicx}
\usepackage{amsmath}
\usepackage{tabularx,booktabs,subcaption}
\usepackage{hyperref}
\newcolumntype{Y}{>{\centering\arraybackslash}X}
\newcolumntype{C}[1]{>{\centering\let\newline\\\arraybackslash\hspace{0pt}}m{#1}}
\usepackage[dvipsnames]{xcolor}
\begin{document}
\title{Generalized Wasserstein Dice Loss, Test-time Augmentation, and Transformers for the BraTS 2021 challenge}
\titlerunning{GWDL, TTA, and Transformers for the BraTS 2021 challenge}
%
\author{
Lucas Fidon\inst{1}$^{*}$ 
\and
Suprosanna Shit\inst{2}$^{*}$
\and
Ivan Ezhov\inst{2}
\and 
Johannes C. Paetzold\inst{2}
\and 
S\'ebastien Ourselin\inst{1}
\and
Tom Vercauteren\inst{1} 
}
\authorrunning{L. Fidon \& S. Shit et al.}
%
\institute{
School of Biomedical Engineering \& Imaging Sciences, King's College London, UK
\and
Department of Informatics, Technical University of Munich, Germany
}
\maketitle              
\begin{abstract}
Brain tumor segmentation from multiple Magnetic Resonance Imaging (MRI) modalities is a challenging task in medical image computation. The main challenges lie in the generalizability to a variety of scanners and imaging protocols. In this paper, we explore strategies to increase model robustness without increasing inference time. Towards this aim, we explore finding a robust ensemble from models trained using different losses, optimizers, and train-validation data split. Importantly, we explore the inclusion of a transformer in the bottleneck of the U-Net architecture. While we find transformer in the bottleneck performs slightly worse than the baseline U-Net in average, the generalized Wasserstein Dice loss consistently produces superior results. Further, we adopt an efficient test time augmentation strategy for faster and robust inference. Our final ensemble of seven 3D U-Nets with test-time augmentation produces an average dice score of 89.4\% and an average Hausdorff 95\% distance of 10.0 mm when evaluated on the BraTS 2021 testing dataset.
Our code and trained models are publicly available at \url{https://github.com/LucasFidon/TRABIT\_BraTS2021}.

\end{abstract}
\section{Introduction}
Gliomas are the most common malignant brain tumors. Broadly, Gliomas are categorized into aggressive high-grade and slow-growing low-grade types. In both types of Gliomas, changes in tissues caused by tumor cells can be captured using multi-modality Magnetic Resonance Imaging (MRI). The commonly used modalities are T1, T2, contrast-enhanced T1 (ceT1), and FLAIR. These modalities are the default choice for the radiologist to identify the tumor type and its progression stage. Towards this objective, accurate and automatic brain tumor segmentation based on multi-parametric MRI is an active field of research \cite{kofler2020brats} and could support diagnosis, surgery planning \cite{ezhov2019neural,ezhov2020geometry}, follow-up, and radiation therapy~\cite{andres2019po,andres2020dosimetry}. The BraTS 2021 challenge has offered an unique and unprecedented opportunity to machine learning researchers to develop a clinically deployable solution for Glioma multi-class segmentation. 

Aiming for computational efficiency, we use 3D U-Net, and its recent transformer variation, TransUNet \cite{chen2021transunet}, as the primary models and focus on finding better learning schemes, such as, augmentation, loss function, optimizer, and efficient inference routine for ensemble model. Recently it has been shown that different loss function combinations may have a crucial impact on the resultant segmentation \cite{kofler2021we}. In our settings, we use the Generalized Wasserstein Dice loss~\cite{fidon2017generalised} that has shown superior segmentation performance as compared to the mean Dice loss~\cite{milletari2016v,sudre2017generalised,li2017compactness} in the BraTS 2020 challenge~\cite{fidon2020brats} and for other medical image segmentation tasks~\cite{tilborghs2020comparative,blanc2020prognostic}. We investigate the effect of different state-of-the-art optimizers, such as, SGD, SGDP \cite{heo2020adamp}, ASAM \cite{kwon2021asam}. Lastly, we use an efficient test-time ensemble approach for the final segmentation result. 
\section{Methods and Materials}

\begin{figure}[tp!]
    \centering
    \includegraphics[width=\linewidth]{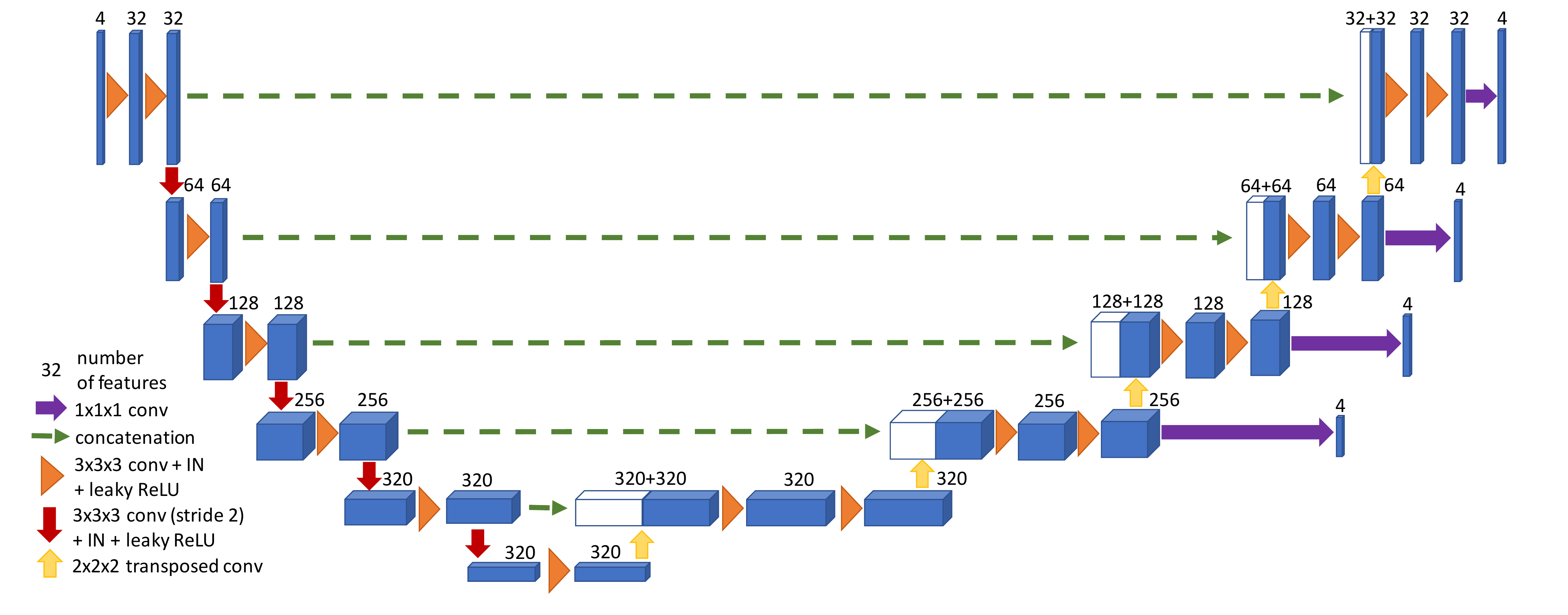}
    \caption{
    \textbf{Illustration of the 3D U-Net~\cite{cciccek20163d} architecture used.}
    Blue boxes represent feature maps. IN stands for instance normalization~\cite{ulyanov2016instance}.
    The design of this 3D U-Net was determined using the heuristics of nnU-Net and our previous work~\cite{fidon2021distributionally,fidon2020brats,fidon2020sgd,isensee2020automated}.
    }
    \label{fig:3dunet}
\end{figure}

\subsection{Data}
We have used the BraTS 2021 dataset\footnote{\url{https://www.synapse.org/\#!Synapse:syn25829067/wiki/610865}}~\cite{baid2021rsna} in our experiments.
No additional data were used.
The dataset contains the same four MRI sequences (T1, ceT1, T2, and FLAIR) for all cases, corresponding to patients with either a high-grade Gliomas~\cite{bakas2017HGG} or a low-grade Gliomas~\cite{bakas2017LGG}.
All the cases were manually segmented for peritumoral edema, enhancing tumor, and non-enhancing tumor core using the same labeling protocol~\cite{menze2014multimodal,bakas2018identifying,bakas2017advancing,baid2021rsna}.
The training dataset contains $1251$ cases, and the validation dataset contains $219$ cases.
MRI for training and validation datasets are publicly available, but only the manual segmentations for the training dataset are available.
The evaluation on the validation dataset was performed using the BraTS 2021 challenge online evaluation platform\footnote{\url{https://www.synapse.org/\#!Synapse:syn25829067/wiki/}}.
For each case, the four MRI sequences are available after co-registration to the same anatomical template, interpolation to $1$mm isotropic resolution, and skull stripping~\cite{menze2014multimodal}.

\subsection{Deep learning pipeline}
We used the DynU-Net of MONAI~\cite{monai} to implement a baseline 3D U-Net with one input block, $4$ down-sampling blocks, one bottleneck block, $5$ upsampling blocks, $32$ features in the first level, instance normalization~\cite{ulyanov2016instance}, and leaky-ReLU with slope $0.01$.
An illustration of the architecture is provided in Fig.~\ref{fig:3dunet}.
We have used the same pipeline for our participation to the FeTA challenge 2021~\cite{fidon2021partial}.

Transformers have recently received attention in medical image computing for their multi-hop attention mechanism. As a second network architecture, we replace the bottleneck block of the U-Net with a vision transformer as proposed by \cite{chen2021transunet}. We use the identical transformer architecture for our experiment as in \cite{chen2021transunet}. A transformer in the bottleneck allows to accumulate the global context of the image and learn an anatomically consistent representation of the tumor classes.
\begin{table}[h]
	\centering
	\caption{Network architecture specification
	}\label{tab:net_arch}
	\begin{tabularx}{0.62\textwidth}{c|c|c}
    	\hline
    	\hline
    	Network & No. of Parameter & Avg. Inference time\\
    	\hline
    	\hline
    	3D U-Net~\cite{cciccek20163d}& 31.2M& 6 seconds\\
    	\hline
    	TransUNet~\cite{chen2021transunet}& 116.7M & 10 seconds\\
    	\hline
    	\hline
	\end{tabularx}
	
\end{table}

Table \ref{tab:net_arch} shows a comparison in terms of the number of parameters and inference time between 3D U-Net and transUNet. For both networks, we train using a patch size of $128 \times 192 \times 128$.

\subsection{Loss function}
We have experimented with two loss functions:
the sum of the cross-entropy loss and the mean-class Dice loss
\begin{equation}
    \mathcal{L}_{DL+CE} = \mathcal{L}_{DL} + \mathcal{L}_{CE}
\end{equation}
and the sum of the cross entropy loss and of the generalized Wasserstein Dice loss\footnote{\url{https://github.com/LucasFidon/GeneralizedWassersteinDiceLoss}}~\cite{fidon2017generalised,fidon2020brats}.
\begin{equation}
    \mathcal{L}_{GWDL+CE} = \mathcal{L}_{GWDL} + \mathcal{L}_{CE}
\end{equation}
where $\mathcal{L}_{CE}$ is the cross entropy loss function
\begin{equation}
    \mathcal{L}_{CE}(\hat{\textbf{p}}, \textbf{p}) = 
    - \sum_{i=1}^N \sum_{l=1}^{L} p_{i,l} \log(\hat{p}_{i,l})
\end{equation}
with $N$ the number of voxels, $L$ the number of classes, $i$ the index for voxels, $l$ the index for classes, $\hat{\textbf{p}}=\left(\hat{p}_{i,l}\right)_{i,l}$ the predicted probability map, and $\textbf{p}=\left(p_{i,l}\right)_{i,l}$ the discrete ground-truth probability map.

$\mathcal{L}_{DL}$ is the mean-class Dice loss~\cite{li2017compactness,sudre2017generalised}
\begin{equation}
    \mathcal{L}_{DL}(\hat{\textbf{p}}, \textbf{p}) = 
    1 - \frac{1}{L}\sum_{l=1}^{L}
    \frac{2 \sum_{i=1}^N  p_{i,l} \hat{p}_{i,l}}{\sum_{i=1}^N  p_{i,l} + \sum_{i=1}^N  \hat{p}_{i,l}}
\end{equation}
And $\mathcal{L}_{GWDL}$ is the generalized Wasserstein Dice loss~\cite{fidon2017generalised}
\begin{equation}
    \left\{
    \begin{split}
        \mathcal{L}_{GWDL}(\hat{\textbf{p}}, \textbf{p}) &= 1 -
        \frac{ 2\sum_{l \neq b} \sum_{i} \textbf{p}_{i,l}(1 - W^M(\hat{\textbf{p}}_i, \textbf{p}_{i}))}{2\sum_{l \neq b}[ \sum_{i} p_{i,l}(1 - W^M(\hat{\textbf{p}}_i, \textbf{p}_{i})) ] + \sum_{i} W^M(\hat{\textbf{p}}_i, \textbf{p}_{i})}\\
        \forall i,\quad W^M\left(\hat{\textbf{p}}_i, \textbf{p}_i\right) &=
        \sum_{l=1}^L p_{i,l}\left(\sum_{l'=1}^L M_{l,l'}\hat{p}_{i,l'}\right)
    \end{split}
    \right.
\end{equation}
where $W^M\left(\hat{\textbf{p}}_i, \textbf{p}_i\right)$ is the Wasserstein distance between predicted $\hat{\textbf{p}_i}$ and ground truth $\textbf{p}_i$ discrete probability distribution at voxel i. 
$M= \left(M_{l,l'}\right)_{1 \leq l,\,l' \leq L}$ is a distances matrix between the BraTS 2021 labels, and $b$ is the class number corresponding to the background.
For the classes indices 0: \textit{background}, 1: \textit{enhancing tumor}, 2: \textit{edema}, 3: \textit{non-enhancing tumor}, we set
\begin{equation}
    M = 
    \left(
    \begin{array}{cccc}
         0 & 1   & 1   & 1 \\
         1 & 0   & 0.7 & 0.5 \\
         1 & 0.7 & 0   & 0.6 \\
         1 & 0.5 & 0.6 & 0 \\
    \end{array}{}
    \right)
\end{equation}

The generalized Wasserstein Dice loss~\cite{fidon2017generalised} is a generalization of the Dice Loss for multi-class segmentation that can take advantage of the hierarchical structure of the set of classes in BraTS.
When the labeling of a voxel is ambiguous or too difficult for the neural network to predict it correctly, the generalized Wasserstein Dice loss and our matrix $M$ are designed to favor mistakes that remain consistent with the sub-regions used in the evaluation of BraTS, i.e., core tumor and whole tumor.

\subsection{Optimization}
\subsubsection{Common optimization setting:}
For each network, the training dataset was split into $95\%$ training and $5\%$ validation at random.
The random initialization of the weights was performed using He initialization~\cite{he2015delving} for all the deep neural network architectures.
We used batch size $2$.
The CNN parameters used at inference corresponds to the last epoch.
We used deep supervision with $4$ levels during training.
Training each 3D U-Net required $16$GB of GPU memory.

\subsubsection{SGD:}
SGD with Nesterov momentum.
The initial learning rate was $0.02$, and we used polynomial learning rate decay with power $0.9$ for a total of $500$ epochs.

\subsubsection{ADAM~\cite{kingma2014adam}:}
For Adam, we used a linear warmup for $1000$ iterations for the learning rate from $0$ to $0.003$ followed by a constant learning rate schedule at the value $0.003$ for $500$ epochs.

\subsubsection{Adaptive Sharpness-Aware Minimization (ASAM)~\cite{kwon2021asam,foret2020sharpness}:}
We have used SGD as the base optimizer with the initial learning rate set to $0.02$ and we used polynomial learning rate decay with power $0.9$ for a total of $500$ epochs.
We used the default hyperparameters of ASAM~\cite{kwon2021asam}, $\rho=0.5$, and $\eta=0.1$.
We have used the PyTorch implementation of the authors\footnote{\url{https://github.com/SamsungLabs/ASAM}}.

\subsubsection{SGDP~\cite{heo2020adamp}:}
For SGD Projected (SGDP), we have used the exact same hyperparameter values as for SGD.
We have used the PyTorch implementation of the authors\footnote{\url{https://github.com/clovaai/AdamP}}.

\subsection{Data augmentation}
We have used random zoom (zoom ratio range $[0.7, 1.5]$ drawn uniformly at random; probability of augmentation $0.3$),
random rotation (rotation angle range $[-15^{\circ}, 15^{\circ}]$ for all dimensions drawn uniformly at random; probability of augmentation $0.3$),
random additive Gaussian noise (mean $0$, standard deviation $0.1$; probability of augmentation $0.3$),
random Gaussian spatial smoothing (standard deviation range $[0.5, 1.5]$ in voxels for all dimensions drawn uniformly at random; probability of augmentation $0.2$),
random gamma augmentation (gamma range $[0.7, 1.5]$ drawn uniformly at random; probability of augmentation $0.3$),
and
random right/left flip (probability of augmentation $0.5$).

\subsection{Inference}
\subsubsection{Single models inference:}
For the models evaluated and compared in Figure~\ref{tab:results}, a patch-based approach is used.
The input image is divided into overlapping patches of size $128 \times 192 \times 128$. The patches are chosen, so that neighboring patches have an overlap of at least half of their volume.
The fusion of the patch prediction is performed using a weighted average of the patch predictions before the softmax operation.
The weights are defined with respect to the distance of a voxel to the center of the patch using a Gaussian kernel standard deviation for each dimension equal to $0.125 \times \textup{patch-dimension}$.
In addition, test-time augmentation~\cite{wang2019aleatoric} is used with right-left flip.
The two softmax predictions obtained with and without right-left flip are merged by averaging.

\subsubsection{Ensemble inference:}
For the ensembles, the inference is performed in two steps.
During the first step, a first segmentation is computed using only one model and the inference procedure for single models. In practice, we used the first model of the list and did not tune the choice of this model.

The first segmentation is used to estimate the center of gravity of the whole tumor.
In the second step, we crop a patch of size $128 \times 192 \times 128$ with a center chosen as close as possible to the center of gravity of the tumor so that the patch fits in the image.
The segmentation probability predictions of all the models of the ensemble are computed for this patch.
The motivation for this two-step approach is to reduce the inference time as compared to using the patch-based approach described above for all the models of the ensemble.
This strategy is based on the assumption that a patch of size $128 \times 192 \times 128$ is large enough to always contain the whole tumor.
During the second step, test-time augmentations with right-left flip and zoom with a ratio of $1.125$ are used.
The four segmentation probability predictions obtained for the different augmentations (no flip - no zoom, flip - no zoom, no flip - zoom, and flip - zoom) are combined by averaging the softmax predictions.
For the full image segmentation prediction, the voxels outside the patch centered on the tumor are set to the background.

\section{Results}
As a primary metric, we report the mean and the standard deviation of the Dice score and the Hausdorff distance for each class. Percentiles are common statistics for measuring the robustness of automatic segmentations~\cite{fidon2021distributionally}.
To evaluate the robustness of the different models, we report the percentiles of the Dice score at $25\%$ and $5\%$ and the percentiles at $75\%$ and $95\%$ of the Hausdorff 95\% distance.
In Table \ref{tab:results}, we report the validation scores of our individually trained models. In Table \ref{tab:results_ensemble} we compare two ensemble strategies as described in the previous section. In the ensemble models, we don't include the TransUNet model as their individual performance is marginally worse than the 3D U-Net model.
\begin{table}[tp]
	\centering
	\caption{\textbf{Segmentation results on the BraTS 2021 Validation dataset.}
	The evaluation was performed on the BraTS online evaluation platform.
	ET: Enhancing Tumor,
	WT: Whole Tumor,
	TC: Tumor Core,
	Std: Standard deviation,
	$p_x$: Percentile x.
	The split number corresponds to the random seed that was used to split the training dataset into $95\%$ training / $5\%$ validation at random.
	}\label{tab:results}
	\begin{tabularx}{\textwidth}{c c *{8}{Y}}
		\toprule
        \multicolumn{2}{c}{}
        & \multicolumn{4}{c}{Dice Score ($\%$)} & \multicolumn{4}{c}{Hausdorff $95\%$ (mm)}\\
        \cmidrule(lr){3-6} \cmidrule(lr){7-10}
		\multicolumn{1}{c}{\bf Model} 
		& \multicolumn{1}{c}{\bf ROI} 
		& Mean & Std & $p_{25}$ & $p_5$
		& Mean & Std & $p_{75}$ & $p_{95}$\\ 
	\midrule
	3D U-Net
		& ET 
		& 82.6 & 23.7 & 83.6 & 7.7
		& 17.9 & 73.7 & 2.2 & 25.4\\
	GWDL + CE SGD
		& TC 
		& 86.4 & 20.1 & 86.8 & 37.7 
		& 11.2 & 50.1 & 4.2 & 19.4\\
	split 1
		& WT
		& 92.5 & 7.4 & 90.7 & 82.1
		& 3.8 & 5.7 & 3.6 & 17.7\\
	\cmidrule(lr){1-10}
	3D U-Net 
		& ET 
		& 82.2 & 24.0 & 82.9 & 7.2
		& 17.8 & 73.7 & 2.4 & 18.1\\
	GWDL + CE SGD
		& TC 
		& 86.5 & 20.4 & 86.1 & 37.7 
		& 11.1 & 50.1 & 4.4 & 21.3\\
	split 2
		& WT
		& 92.5 & 7.4 & 90.6 & 82.6
		& 3.8 & 5.9 & 3.7 & 11.5\\
	\cmidrule(lr){1-10}
	3D U-Net 
		& ET 
		& 81.9 & 24.5 & 82.5 & 5.3
		& 19.5 & 77.5 & 2.2 & 31.7\\
	GWDL + CE SGD
		& TC 
		& 85.7 & 21.6 & 86.0 & 30.5 
		& 11.5 & 50.1 & 4.2 & 20.8\\
	split 27
		& WT
		& 92.5 & 7.1 & 90.5 & 81.7
		& 4.0 & 6.2 & 3.9 & 10.6\\
	\cmidrule(lr){1-10}
	3D U-Net 
		& ET 
		& 81.9 & 24.9 & 83.0 & 0.0
		& 21.1 & 81.0 & 2.2 & 67.0\\
	GWDL + CE SGD
		& TC 
		& 86.5 & 20.7 & 87.2 & 38.5 
		& 9.5 & 43.7 & 4.1 & 16.9\\
	split 1227
		& WT
		& 92.5 & 7.3 & 90.6 & 81.1
		& 3.8 & 5.8 & 3.7 & 11.7\\
	\cmidrule(lr){1-10}
	3D U-Net 
		& ET 
		& 82.4 & 24.4 & 83.2 & 2.9
		& 19.5 & 77.5 & 2.4 & 30.3\\
	GWDL + CE SGD
		& TC 
		& 85.8 & 21.7 & 86.3 & 27.6 
		& 11.5 & 50.1 & 4.6 & 23.0\\
	split 122712
		& WT
		& 92.4 & 7.1 & 90.2 & 81.7
		& 4.1 & 7.0 & 3.9 & 11.4\\
	\cmidrule(lr){1-10}
	3D U-Net
		& ET 
		& 81.7 & 24.9 & 82.9 & 0.0
		& 21.3 & 81.1 & 2.4 & 89.3\\
	DL + CE SGD
		& TC 
		& 86.5 & 20.4 & 86.3 & 40.2 
		& 11.0 & 50.1 & 4.1 & 17.9\\
	split 1
		& WT
		& 92.5 & 7.2 & 90.6 & 80.2
		& 3.9 & 6.7 & 3.9 & 9.8\\
	\cmidrule(lr){1-10}
	3D U-Net 
		& ET 
		& 82.1 & 24.2 & 82.8 & 5.4
		& 19.5 & 77.5 & 2.5 & 31.3\\
	GWDL + CE SGDP
		& TC 
		& 86.3 & 20.5 & 86.4 & 40.4 
		& 9.7 & 43.7 & 4.1 & 20.2\\
	split 1
		& WT
		& 92.6 & 7.4 & 90.5 & 82.0
		& 3.8 & 5.9 & 3.7 & 11.0\\
	\cmidrule(lr){1-10}
	3D U-Net
		& ET 
		& 80.0 & 25.7 & 81.5 & 0.0
		& 23.4 & 84.5 & 3.0 & 373.1\\
	GWDL + CE ASAM
		& TC 
		& 85.4 & 21.5 & 86.1 & 35.3 
		& 12.2 & 50.8 & 4.2 & 21.7\\
	split 1
		& WT
		& 91.9 & 7.9 & 90.0 & 77.2
		& 5.4 & 10.7 & 4.2 & 22.0\\
	\cmidrule(lr){1-10}
	TransUNet 
	    & ET 
		& 79.7 & 25.3 & 80.1 & 0.0
		& 22.4 & 81.0 & 3.0 & 113.4\\
		GWDL + CE SGD & TC 
		& 84.5 & 22.0 & 84.5 & 36.5 
		& 8.7 & 36.5 & 4.9 & 24.9\\
		split 1 & WT
		& 91.8 & 7.3 & 90.2 & 77.2
		& 4.4 & 7.6 & 4.1 & 12.7\\
	\cmidrule(lr){1-10}
	TransUNet 
	    & ET 
		& 80.9 & 24.1 & 81.2 & 6.5
		& 18.5 & 73.6 & 3.0 & 37.7\\
		GWDL + CE ADAM & TC 
		& 84.8 & 21.9 & 84.7 & 36.44 
		& 10.3 & 43.8 & 4.6 & 24.6\\
		split 1 & WT
		& 91.9 & 7.8 & 90.0 & 80.1
		& 4.1 & 6.5 & 4.1 & 14.9\\
	\cmidrule(lr){1-10}
	TransUNet
	    & ET 
		& 80.2 & 25.5 & 80.4 & 0.0
		& 23.1 & 84.3 & 3.0 & 373.1\\
		GWDL + CE SGDP & TC 
		& 85.2 & 21.9 & 85.6 & 27.6 
		& 11.45 & 50.0 & 4.5 & 19.9\\
		split 1 & WT
		& 92.0 & 7.4 & 90.1 & 79.7
		& 4.7 & 9.1 & 4.0 & 18.0\\
	\bottomrule
	\end{tabularx}
\end{table}

\begin{table}[tp]
	\centering
	\caption{\textbf{Segmentation results on the BraTS 2021 Validation dataset for ensembling and test-time augmentation.}
	The evaluation was performed on the BraTS online evaluation platform.
	ET: Enhancing Tumor,
	WT: Whole Tumor,
	TC: Tumor Core,
	Std: Standard deviation,
	$p_x$: Percentile x.
	Best values are in \textbf{bold}.
	}\label{tab:results_ensemble}
	\begin{tabularx}{\textwidth}{c c *{8}{Y}}
		\toprule
        \multicolumn{2}{c}{}
        & \multicolumn{4}{c}{Dice Score ($\%$)} & \multicolumn{4}{c}{Hausdorff $95\%$ (mm)}\\
        \cmidrule(lr){3-6} \cmidrule(lr){7-10}
		\multicolumn{1}{c}{\bf Model} 
		& \multicolumn{1}{c}{\bf ROI} 
		& Mean & Std & $p_{25}$ & $p_5$
		& Mean & Std & $p_{75}$ & $p_{95}$\\ 
	\midrule
	3D U-Net
		& ET 
		& 82.0 & 24.4 & 83.0 & 3.1
		& 19.5 & 77.5 & 2.3 & 30.0\\
	ensemble
		& TC 
		& 86.6 & 20.2 & 86.5 & 39.2 
		& 9.5 & 43.7 & 4.1 & 20.1\\
		& WT
		& 92.6 & 7.2 & 90.6 & 82.0
		& 3.9 & 6.3 & 3.6 & 12.8\\
	\cmidrule(lr){1-10}
	3D U-Net
		& ET 
		& \bf 84.0 & \bf 22.0 & \bf 84.2 & \bf 21.6
		& \bf 12.7 & \bf 60.6 & \bf 2.2 & \bf 16.0\\
	ensemble
		& TC 
		& \bf 87.0 & \bf 20.0 & \bf 86.8 & \bf 43.1 
		& \bf 11.0 & \bf 50.1 & \bf 4.1 & \bf 18.7\\
	zoom augmentation
		& WT
		& \bf 92.7 & \bf 7.2 & \bf 90.6 & \bf 82.1
		& \bf 3.9 & \bf 6.3 & \bf 3.6 & \bf 12.6\\
	\bottomrule
	\end{tabularx}
\end{table}

\begin{table}[tp]
	\centering
	\caption{\textbf{Segmentation results on the BraTS 2021 Testing dataset using ensembling and test-time augmentation.}
	The evaluation was performed by the BraTS 2021 challenge organizers using our docker submission.
	ET: Enhancing Tumor,
	WT: Whole Tumor,
	TC: Tumor Core,
	Std: Standard deviation,
	$p_x$: Percentile x.
	}\label{tab:results_ensemble_test}
	\begin{tabularx}{\textwidth}{c c *{6}{Y}}
		\toprule
        \multicolumn{2}{c}{}
        & \multicolumn{3}{c}{Dice Score ($\%$)} & \multicolumn{3}{c}{Hausdorff $95\%$ (mm)}\\
        \cmidrule(lr){3-5} \cmidrule(lr){6-8}
		\multicolumn{1}{c}{\bf Model} 
		& \multicolumn{1}{c}{\bf ROI} 
		& Mean & Std & $p_{25}$
		& Mean & Std & $p_{75}$\\ 
	\midrule
	3D U-Net
		& ET 
		& 87.4 & 17.6 & 85.2 
		& 10.1 & 53.5 & 2.0 \\
	ensemble
		& TC 
		& 87.8 & 23.6 & 91.3
		& 15.8 & 66.7 & 3.0 \\
	zoom augmentation
		& WT
		& 92.9 & 9.0 & 91.6
		& 4.1 & 7.3 & 3.7\\
	\bottomrule
	\end{tabularx}
\end{table}

\section{Discussion}

From Table \ref{tab:results}, we see that 3D U-Net trained with generalized Wasserstein Dice loss performs consistently better than the one with Dice loss (baseline model). TransUNet does not offer any improvement over the baseline. Rather the performance deteriorates slightly. We hypothesize that over-parameterization can be an issue in this case. The optimizer SGDP and ASAM perform similar to the baseline SGD. From Table \ref{tab:results_ensemble}, we see that ensemble strategy helps in increasing the robustness of the model. The best ensemble strategy turns out to be including zoom as a test time augmentation.
This approach was submitted for evaluation on the BraTS 2021 testing dataset and the results can be found in Table \ref{tab:results_ensemble_test}.
In conclusion, this paper proposes a detailed comparative study on the strategies to make a computationally efficient yet robust automatic brain tumor segmentation model. We have explored ensemble from multiple training configurations of different state-of-the-art loss functions and optimizers, and importantly, test-time augmentation. Future research will focus on further strategies on test-time augmentation and test-time hyper-parameter tuning.

\subsubsection*{Acknowledgments}
This project has received funding from the European Union's Horizon 2020 research and innovation program under the Marie Sk{\l}odowska-Curie grant agreement TRABIT No 765148;
Wellcome [203148/Z/16/Z; WT101957], EPSRC [NS/A000049/1; NS/A000027/1].
Tom Vercauteren is supported by a Medtronic / RAEng Research Chair [RCSRF1819\textbackslash7\textbackslash34].
Data used in this publication were obtained as part of the RSNA-ASNR-MICCAI Brain Tumor Segmentation (BraTS) Challenge project through Synapse ID (syn25829067).

%
%
%
\bibliographystyle{splncs04}
\bibliography{main}

\end{document}